%% file: spec.tex
\documentclass[twoside]{sfm}
\input{sf.def}
\usepackage[usenames,dvipsnames]{xcolor}

\def\llm{{\sc LLmodels}}

\def\inv{{\sc invers10}}

\newcommand{\bz}{$\langle B_\mathrm{z} \rangle$}
\newcommand{\ion}[2]{#1{\sc #2}}

\begin{document}
\input{rusomarov/rusomarov.tex}

\end{document}

%% file: rusomarov/rusomarov.tex
\pagebreak

\thispagestyle{titlehead}

\setcounter{section}{0}
\setcounter{figure}{0}
\setcounter{table}{0}

\markboth{Rusomarov et al.}{Ap stars from four Stokes parameters}{}

\titl{Magnetic fields of Ap stars from the full Stokes spectropolarimetric observations}{Rusomarov N., Kochukhov O. \& Piskunov N.}
{Division of Astronomy and Space Physics, Department of  Physics and Astronomy, Uppsala University, Sweden\\ email: {\tt naum.rusomarov@physics.uu.se}}

\abstre{Current knowledge about stellar magnetic fields relies almost entirely on circular polarization observations. Few objects have been observed in all four Stokes parameters. The magnetic Ap star HD\,24712 (DO Eri, HR\,1217) was recently observed in the Stokes~$IQUV$ parameters with the HARPSpol instrument at the 3.6-m ESO telescope as part of our project at investigating Ap stars in all four Stokes parameters. The resulting spectra have dense phase coverage, resolving power $> 10^5$, and S/N ratio of 300--600. These are the highest quality full Stokes observations obtained for any star other than the Sun.

We present preliminary results from magnetic Doppler imaging of HD\,24712. This analysis is the first step towards obtaining detailed 3-D maps of magnetic fields and abundance structures for HD\,24712 and other Ap stars that we currently observe with HARPSpol.}

\baselineskip 12pt

\section{Introduction}
Magnetic Ap stars are interesting objects that have been investigated in many ways. One aspect of these investigations is the interplay between their magnetic fields and chemical spots. However, such studies as a rule use only circular polarization observations, and in rare cases additional broadband linear polarization measurements. Latest spectropolarimetric investigations that have used four Stokes parameter observations show that these stars posses strong magnetic fields that are dipolar at large spatial scales with significant complexity at small scales \cite{Kochukhov10p13}. Furthermore, these investigations show that magnetic Ap stars have inhomogeneous abundance distribution of many chemical elements \cite{Kochukhov2004p613a}. Other spectroscopic and photometric pulsational analysis studies \cite{Shulyak09p879,Ryabchikova2007p1103} show that some chemically peculiar stars have strong abundance gradients in their atmospheres. 

Recent advances in observational techniques in spectropolarimetry are capable of bringing new insight into the problem of magnetic Ap stars. The spectropolarimeter mode for the HARPS spectrograph \cite{Mayor2003p20} at the 3.6-m ESO telescope allows for \textit{simultaneous} observations in all four Stokes parameters with spectral resolution greater than $10^5$. Unfortunately, current model atmospheres and imaging techniques, e.g., \llm{} \cite{Shulyak2004p993} and magnetic Doppler imaging (MDI) \cite{Piskunov2002p736}, although very powerful in itself, have various limitations, which ultimately prevent us from exploiting the new advances in observational techniques. MDI codes do not incorporate vertical chemical stratification, while empirical stratified model atmospheres do not consider horizontal variations due to stellar surface structure.

We believe that it is possible to lift these limitations on the current methods and perform simultaneous study of 3-D chemical and magnetic structures using highest quality four Stokes parameter observations. Such investigations can give us more insight into the relation between chemical spots and magnetic field, and between horizontal and vertical structures, and will allow us to use the exceptional observations to the fullest extent.

\section{Spectropolarimetric observations}
\begin{figure}[t]
  \centering
  \includegraphics[width=\textwidth]{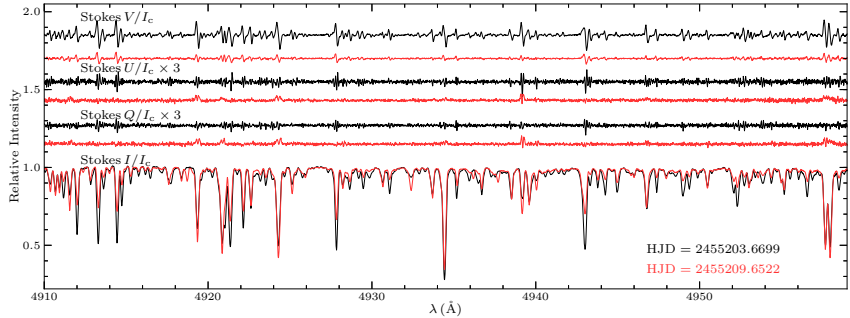}
  \caption{HARPSpol four Stokes parameter spectra of HD\,24712. The spectra plotted with black lines were obtained around zero phase during magnetic maximum; spectra plotted with red lines were obtained at phase corresponding to magnetic minimum. The different Stokes profiles are offset vertically for clarity. The Stokes~$Q$ and $U$ spectra are expanded by a factor of three compared to Stokes~$V$ and $I$. Notice how most lines exhibit strong intensity variations with phase, and have strong linear and circular polarization signatures.}
  \label{fig:order7}
\end{figure}

For this purpose we have started new program aimed at observing Ap stars in all four Stokes parameters with the HARPSpol spectropolarimeter at the 3.6-m ESO telescope. As our first target we have chosen HD\,24712, which is one of the coolest Ap stars showing both stratification of chemical elements and oblique rotator variations. This star was observed during 2010--2011 for sixteen nights, during which 43 individual Stokes parameter observations were obtained. The resulting spectra have S/N ratio of 300--600 and resolving power exceeding $10^5$. During the last two observation runs for the period 2012--2013 we finished observing two more objects, HD\,125248 and HD\,119419. Each of these objects has 12 Stokes~$IQUV$ observations with S/N ratio $\sim$ 200. In Fig.~\ref{fig:order7} we present observed spectra of HD\,24712 in all four Stokes parameters in the 4910--4960\,\AA{} region. We can see that many individual spectral lines show strong, complex linear and circular polarization signatures. Many lines exhibit significant polarization signatures as part of complex blends. 

An initial analysis of the full Stokes vector spectropolarimetric data set of HD\,24712 has been published by Rusomarov et al. \cite{Rusomarov2013pA8}. In that paper we computed mean line profiles using least-squares deconvolution and determined magnetic observables from low-order moments of these profiles. We measured the mean longitudinal magnetic field, \bz{}, with an accuracy of 5--10\,G and obtained precise net linear polarization measurements. We combined \bz{} and net linear polarization measurements and determined parameters of the dipolar magnetic field topology. Combining available \bz{} measurements we improved the rotational period. The analysis of all \bz{} measurements showed no evidence for a significant radial magnetic field gradient.

\section{Magnetic Doppler Imaging}
Here we present preliminary results from the magnetic Doppler imaging of HD\,24712. For this purpose we used the MDI code \inv{}, first introduced by Piskunov \& Kochukhov \cite{Piskunov2002p736}, which is used for simultaneous and self-consistent reconstruction of the magnetic and chemical maps of the stellar surface using spectra in all four Stokes parameters. The inversion procedure employs synthetic spectra calculated treating in detail polarized radiative transfer in the atmosphere of the star. The code is written in Fortran and is optimized for running on massively parallel computers using MPI libraries.

To carry out the inversion procedure besides the observed profiles and line data we also need to supply a model atmosphere and a number of input parameters. We chose an \llm{} model atmosphere for the stellar parameters presented in Rusomarov et al. \cite{Rusomarov2013pA8}. The abundance values for the model atmosphere of the most important elements (Fe-peak elements and REEs) were set to the values that correspond to the mean spectrum. Among the (constant) parameters that we need to supply to \inv{} are the inclination of the stellar axis, $i$, and the obliquity angle, $\beta$. We adopted the values for these parameters derived by Rusomarov et al. \cite{Rusomarov2013pA8} for the case of a dipolar magnetic field topology. 

\begin{figure}[h!]
  \centering
  \includegraphics[width=0.85\textwidth, height=0.90\textheight, bb=45 75 450 715, clip]{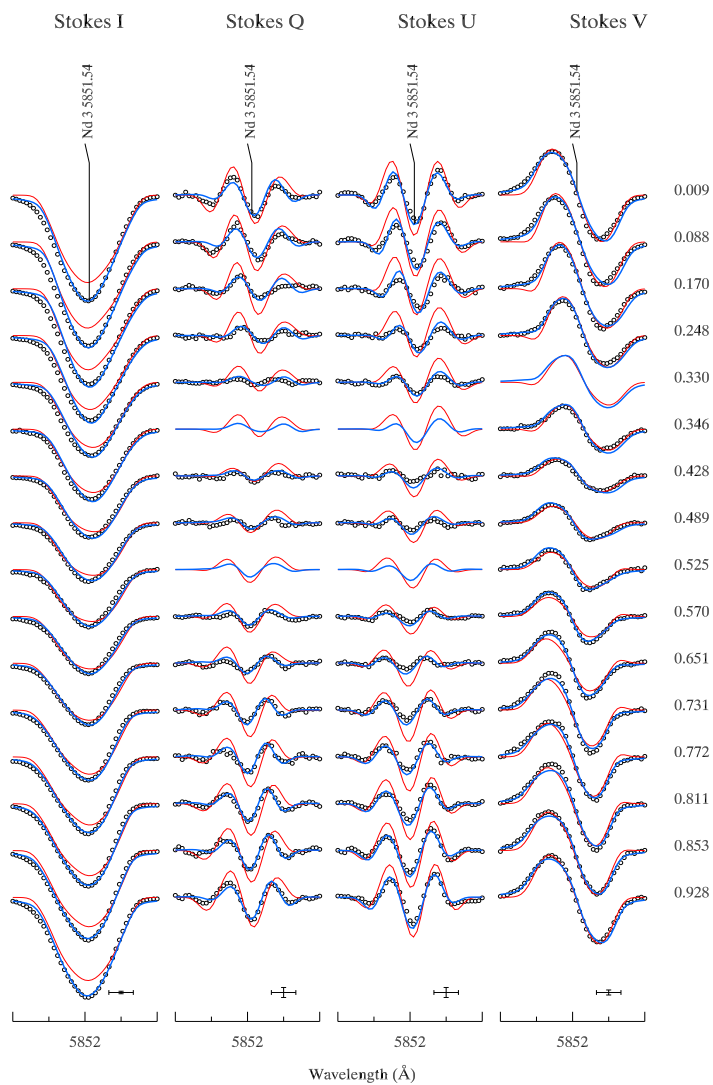}
  \caption{Observed (dots) and calculated (thick lines) Stokes profiles for HD\,24712. With thin lines we plotted the Stokes profiles calculated according to the Nd abundance maps and magnetic field distribution derived by L\"uftinger et al. (2010). Phase values increase downwards. The bars at the bottom of each panel show the vertical (5\,\%) and horizontal (0.5\,\AA{}) scale.}
  \label{fig:MDI_prof}
\end{figure}

Since in this paper we are not aiming for a complete inversion using as many spectral lines as possible we chose the unblended \ion{Nd}{iii} 5851.54\,\AA{} line that shows very strong linear and circular polarization signatures. We set additional constraint in \inv{} that keeps the geometry of the magnetic field dipolar. This is needed because HD\,24712 shows mostly positive magnetic pole to the observer.
\begin{figure}[t]
  \centering
  \includegraphics[width=0.85\textwidth, bb=27 185 570 300, clip]{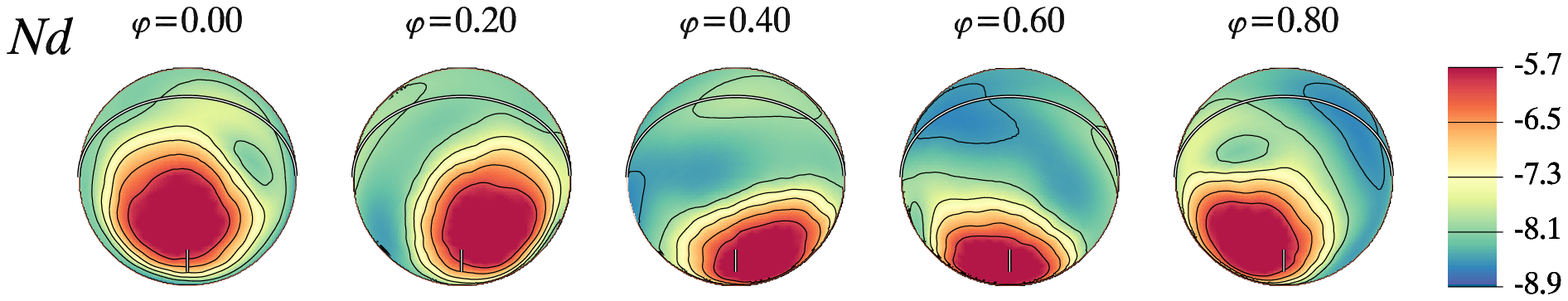}\\
  \includegraphics[width=0.85\textwidth, bb=27 180 570 470, clip]{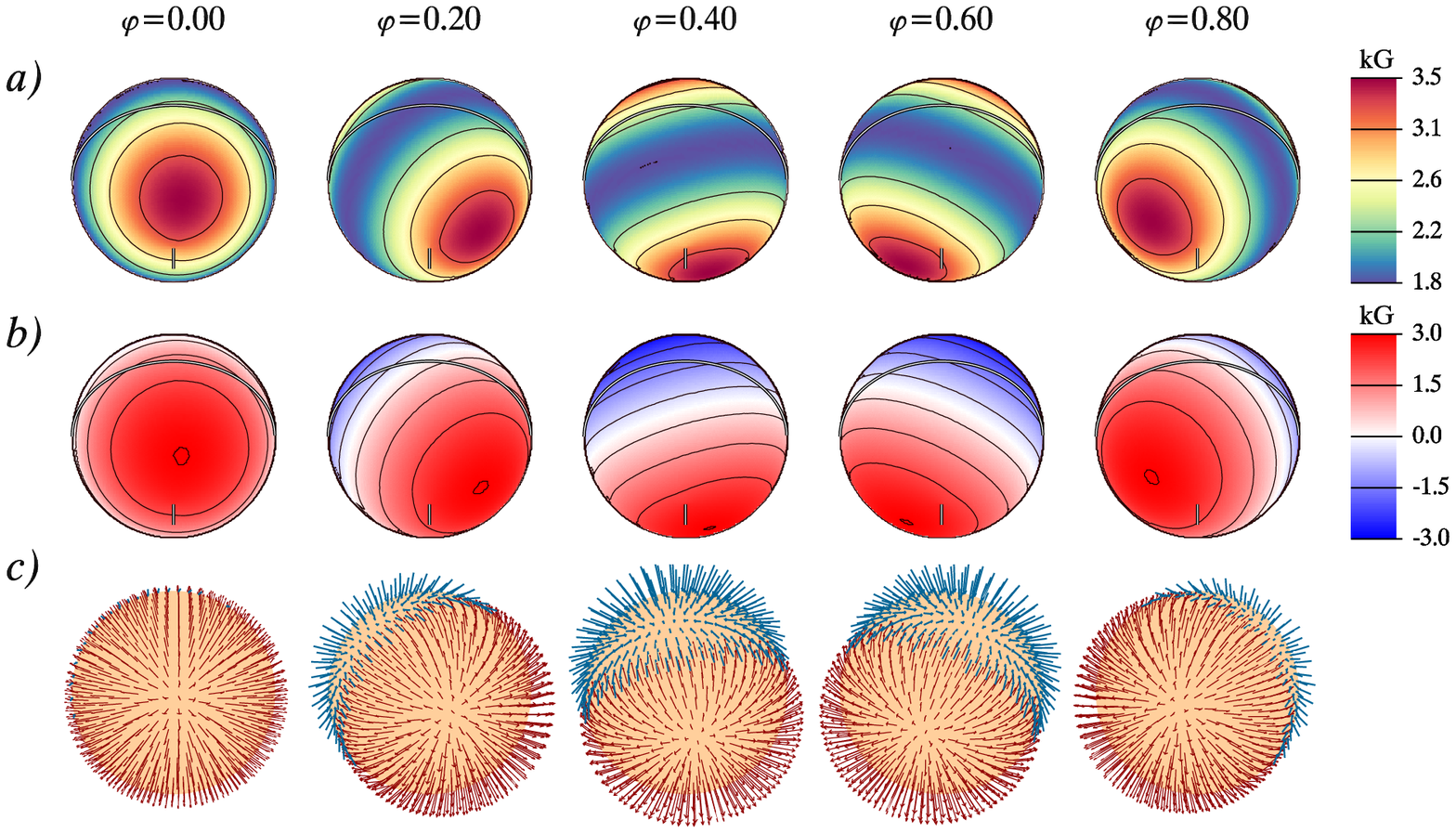}
  \caption{Abundance distribution maps of Nd on the surface of HD24712, and maps showing a distribution of the magnetic field strength a), radial component b), and field orientation c).}
  \label{fig:MDI_ab_mf}
\end{figure}
The inversion results of our MDI analysis are shown in Fig.~\ref{fig:MDI_ab_mf}. From this figure we can see that abundance distribution maps of Nd and maps of the magnetic field are not too far off from the results by L\"uftinger et al. \cite{Luftinger2010p71}. Neodymium is distributed in a patch around the visible magnetic pole. In Fig.~\ref{fig:MDI_prof} we compare observations and model proflies for the \ion{Nd}{iii} line. At this stage one can make an interesting comparison between our results from one line and the results by L\"uftinger et al. \cite{Luftinger2010p71}, who deduced the magnetic field distribution and abundance maps of various elements using only Stokes~$IV$ spectra under the assumption of dipole field geometry. When we compare the profiles of \ion{Nd}{iii} 5851.54\,\AA{} calculated according to their results we see that Stokes~$I$ and $V$ profiles with with good success describe our observations. However, Stokes~$Q$ and $U$ profiles do not match in amplitude, although, they match in morphology. This may be expected since L\"uftinger et al. \cite{Luftinger2010p71} did not use linear polarization data in any way. In any case, our preliminary MDI results do not show evidence of significant deviations from dipolar field geometry found in other MDI four stokes parameter studies of Ap stars \cite{Kochukhov10p13,Kochukhov2004p613a}.

In the future we will perform a more detailed MDI analysis with lines from more chemical elements for HD\,24712. We plan on obtaining detailed maps of the magnetic field and horizontal and vertical abundance structures not only for HD\,24712, but also for other Ap stars.
